\documentclass[a4paper,12pt,BCOR5mm,bibliography=totoc]{article}

\usepackage{caption}
\usepackage{subcaption}

\usepackage[english]{babel}
\usepackage{fancyhdr}
\usepackage{setspace}
\usepackage{graphicx}
\usepackage{eso-pic}
\usepackage{amsmath}
\usepackage{amstext}
\usepackage{amsfonts}
\usepackage{amssymb}
\usepackage{hyperref}
\usepackage{slashed}
\usepackage{cite}
\usepackage{bbold}
\usepackage{multirow}
\usepackage{esdiff}
\usepackage[font=small,labelfont=small]{caption}
\usepackage{commath}
\usepackage{tikz-cd}
\usepackage{mathabx}
\usepackage{relsize}
\usepackage{simplewick} 

\usepackage[affil-it]{authblk} 
 
\newtheorem{theorem}{Theorem}

\renewcommand{\vec}[1]{\ensuremath{\boldsymbol{#1}}}

\newcommand{\N}{\mathbb{N}}	
\newcommand{\Z}{\mathbb{Z}}	
\newcommand{\R}{\mathbb{R}}	
\newcommand{\iu}{\mathrm{i}}	
\newcommand{\ee}{\mathrm{e}}	

\newcommand{\bra}[1]{\ensuremath{\left< #1\,\right|}}
\newcommand{\ket}[1]{\ensuremath{\left|\, #1\right>}}
\newcommand{\braket}[2]{\ensuremath{\left< #1\, |\, #2\right>}}
\newcommand{\ketbra}[2]{\ket{#1}\bra{#2}}

\begin{document}

\vspace{0.01cm}
\begin{center}
{\Large\bf The Lieb-Liniger model at the critical point as toy model for Black Holes} 

\end{center}

\vspace{0.1cm}

\begin{center}

{\bf Mischa Panchenko}$^{a}$\footnote{m.panchenko@campus.lmu.de} 

\vspace{.6truecm}


{\em $^a$Arnold Sommerfeld Center for Theoretical Physics\\
Department f\"ur Physik, Ludwig-Maximilians-Universit\"at M\"unchen\\
Theresienstr.~37, 80333 M\"unchen, Germany}

\end{center}

\vspace{0.5cm}

\begin{abstract}
\noindent  
 
{\small 
In a series of papers \cite{cesargia} it was proposed that black holes can be understood as Bose-Einstein condensates at the critical point of a quantum phase transition. Therefore other bosonic systems with quantum criticalities, such as the Lieb-Liniger model with attractive interactions, could possibly be used as toy models for black holes. Even such simple models are hard to analyse, as mean field theory usually breaks down at the critical point. Very few analytic results are known. In this paper we present a method of studying such systems at quantum critical points analytically. We will be able to find explicit expressions for the low energy spectrum of the Lieb-Liniger model and thereby to confirm the expected black hole like properties of such systems. This opens up an exciting possibility of constructing and studying black hole like systems in the laboratory.
}

\end{abstract}

\thispagestyle{empty}
\clearpage

\section{Introduction and results}

Black holes (BH) have been puzzling physicists ever since they were found within general relativity. In particular their quantum nature and information processing is poorly understood. In a series of papers \cite{cesargia} the authors proposed, that black holes can be understood as leaky Bose-Einstein condensates at the point of a quantum phase transition. While for BH the coupling is self-finetuned, so that the condensate stays critical throughout the evaporation process, we can ask ourselves whether an ordinary system of bosons with a quantum phase transition displays any properties of the BH. This line of thought was initiated in \cite{cesargia2} - \cite{gold}. There it was shown that such systems of attractive bosons indeed do possess some of the key properties of black holes. In the present paper we will perform an analytical study of the proposed system, in order to extract scaling properties, the ground state structure etc. In particular, we confirm the appearance of gapless states analytically, find the scaling of the energy gap and develop a method for studying the dynamics of such systems at the critical point. This method was used and elaborated in \cite{giamischa}. It turns out that the entanglement generation is slow even at the critical coupling, so that the Bogoliubov approximation is valid for long times. In \cite{giamischa} it was then explicitly shown how to encode and decode information using the attractive bosons close to the critical point. This leads to the exciting possibility of building  BH like systems in a laboratory and studying their information processing properties. The advance in the analytical understanding of the phase transition in the attractive Lieb-Liniger model as presented in this paper, will then enable us to make quantitative predictions for such experiments. \\

Concretely, we will use a new diagonalization technique, developed in more detail in \cite{mischa}, to analyze the system of non-relativistic Bosons on a one-dimensional Ring with attractive delta interaction, also known as the Lieb-Liniger model \cite{Lieb}. \\
  
The Hamiltonian of the system can be written as     
   \begin{equation}
 {\mathcal H} \, = \, \int d^x \, \psi^{+} \frac{- \hbar^2 \Delta}{ 2m} \psi \, - \, g \hbar  \ 
  \int dx \, \psi^{+}\psi^{+} \,\psi \psi \,, 
\label{Hnonderivative} 
\end{equation} 
where $\psi \, = \, \sum\limits_{k\in\Z} \frac{1} {\sqrt{V}} {\rm e}^{i  {k \over R} \vec{x}} \, a_k$, 
$V \, = \, 2\pi R$ is the 1-dimensional volume and $k$ the (angular) momentum.  
$a_{k}^\dagger, a_{k}$ are the usual creation and annihilation operators of bosons of momentum-number 
${k}$. These operators satisfy: $[a_{k}, a_{k'}^\dagger] = \delta_{kk'}$
and all other commutators vanish. 
 The parameter $g$ controls the strength of the coupling. 

 We will represent the Hamiltonian in the form  
 $ {\mathcal H} \, \equiv \, {\hbar^2 \over 2R^2 m} \, H$.
  The quantity ${\hbar^2 \over 2R^2 m}$ is a unit for energy measurement, since 
  it gives us an idea about the energy cost of a given process 
  relative to the kinetic energy of the first non-zero momentum mode of a single free boson. 
Therefore all our further discussions will take place in the units ${\hbar^2 \over 2R^2 m} \, = \, 1$. 
   That is, we effectively switch to the Hamiltonian $H$.  

 Introducing a notation $\tilde{\alpha} \, \equiv \left ({g \over V R} \right) {2Rm \over \hbar} $, 
  this Hamiltonian takes the form,
 \begin{equation}
     H \, = \,  \sum_{k}  k^2 \, a_k^\dagger a_k  \, - \, {\tilde{\alpha} \over 4} \, \sum_{k_1+ k_2-k_3-k_4 = 0} 
   a_{k_1}^\dagger a^\dagger _{k_2}  a_{k_3}a_{k_4} \,.   
     \label{Hamilton}
 \end{equation}
 
 The particle number and the total momentum, described by the operators
 \begin{equation}
  \hat{n}=\sum_{k} \, a_k^\dagger a_k \quad,\quad
  P=\sum_{k}  k \, a_k^\dagger a_k,
 \end{equation}
are conserved.\\

The above system exhibits quantum critical behavior - it is known to exhibit a quantum phase transition towards the bright soliton phase, studied in detail in \cite{bright}. This quantum phase transition occurs at $\tilde{\alpha} N=1$ with $N$ being the particle number. From now on, we will restrict ourselves to a fixed $N$-sector with total momentum $P=0$ and rescale the coupling by defining $\alpha:=\tilde{\alpha}N$. The treatment can be easily extended to $P\neq 0$ sectors as well.\\
 
 We will study the system exactly at the critical coupling, where the usual techniques like Bogoliubov diagonalization or mean field methods fail. In the present paper we will demonstrate the following results for the coupling $\alpha=1$ and $P=0$:
 \begin{enumerate}
  \item Only the modes $a_0, a_{\pm1}$ contribute to the low energy part of the spectrum.\label{1}
  \item The energy gap between two neighboring low energy eigenstates scales as $N^{-1/3}$, the system becomes gapless as $N\rightarrow\infty$.\label{2}
  \item A new type of perturbation theory will be developed and applied. An explicit analytic expression for the ground state will be obtained to first order in $\frac{1}{N}$. Correlation functions and higher order corrections can be calculated in a straightforward way.\label{3}
  \item The depletion of the ground state scales with $N^{\frac{1}{3}}$. In the limit $N\rightarrow\infty$ the ground state is completely depleted.\label{4}  
  \item For large $N$, the system is equivalent to a 2-dimensional isotropic purely quartic oscillator. Therefore, the results incidentally reveal several properties of the latter. \label{5}
 \end{enumerate}
All results were obtained analytically and cross-checked with numerical calculations. The paper is structured as follows: we will first restrict ourselves to the  $a_0, a_{\pm1}$ modes and compute the coefficients of the resulting finite dimensional Hamiltonian $H$ in section \ref{section2}. Then we will compute the low energy eigenvectors and eigenstates of $H$ in section \ref{section3} by means of new techniques - this will be the main part of the paper.  Diagonalizing an interacting system is not easy, therefore we split the calculation into several parts. In \ref{subsection31}, after having introduced the necessary notions, we will diagonalize the Bogoliubov part of $H$ before and right at the critical coupling. We will see how the double scaling limit and the breakdown of the Bogoliubov approximation will appear in our language. Then in \ref{subsectionint} we will include the interaction at the critical point and develop a perturbation theory in $1/N$. From the analytic expression for the ground state obtained there, \ref{2} -\ref{4} will follow. We will go on to prove \ref{1} in section \ref{sectionmodes}. Finally we mention the connection to the quartic oscillator in \ref{sectionosc}.

\section{Setting the stage}\label{section2}
When we restrict ourselves to an $N$-particle sector with angular momentum zero and cut off the modes with momenta higher than one, the Hilbert space becomes finite dimensional. Its basis can be written as 
\begin{equation}\label{basis}
  \ket{n}:=\ket{n_{-1}=n,n_0=N-2n,n_{1}=n},                                                                                                                                                                                                                                                                                                                                                                                                           \end{equation}
  
 where $n$ goes from zero to $N/2$. After a simple calculation, one observes that the Hamiltonian in this approximation is a $(\frac{N}{2}+1)\times (\frac{N}{2}+1)$ dimensional tridiagonal matrix of the following form:
\begin{equation}\label{tridiag}
H= \begin{pmatrix}
d_0   & h_{1} & 0  & \dots \\
h_1 &  d_1   & h_{2}  & \ddots \\
0& h_2 & \ddots & \ddots \\
\vdots & \ddots &\ddots &\ddots
\end{pmatrix}
+C_N,
\end{equation}
with the entries 
\begin{equation}
 d_n=(2-\alpha)n+\frac{3\alpha}{2N}n^2 \quad,\quad h_n=-\frac{\alpha}{2}n+\frac{\alpha}{N}n^2
\end{equation}
and $C_N$ is a negative $N$-dependent constant which we will ignore from now on \footnote{ We have approximated $\sqrt{(N-2n)(N-2n-1)}\approx (N-2n)$.}. We will now split the matrix as follows:
\begin{equation}
 H=H_B+H_I \quad \text{ and further } \quad H_I=H_{I_1}+H_{I_2}
\end{equation}
by splitting the coefficients as
\small
\begin{equation}
 d_{B\,n}=(2-\alpha)n \quad,\quad d_{I_1\,n }=\frac{3\alpha}{2N}n^2 \quad,\quad 
  h_{B\,n}=-\frac{\alpha}{2}n  \quad,\quad h_{I_2\,n}=-\frac{\alpha}{N}n^2.
\end{equation}
\normalsize
In other words, $H_B$ is the part of $H$ that is linear in $n$ and $H_I$ is quadratic and suppressed by $1/N$. Then, $H_I$ is further split into the diagonal part $H_{I_1}$ and the off-diagonal rest. The subscripts $B$ and $I$ stand for Bogoliubov and Interaction, since in the Bogoliubov approximation it is exactly $H_B$ which is diagonalized while $H_I$ is ignored.\\

Let us introduce some useful notation. Given a normalized state $\ket{v}$ we can expand it in the above basis to obtain the (finitely many) coefficients $v_n:=\braket{n}{v}$. We will call $\ket{v}$ \textit{localized up to $n_0$} if there exists an $n_0$ s.t. 
\begin{equation}
\forall \, m \in \N \quad v_{n_0+m}\ll 1,
\end{equation}
  i.e. if the coefficients  $v_n$ approximately vanish after some $n_0$ (we will of course take $n_0$ always as small as possible). Notice that for such vectors we find 
  \begin{equation}
  H_{I}\ket{v}=\mathcal{O}\left(\frac{n_0^2}{N}\right).
  \end{equation}
This simple observation is one of the key factors of the following calculations.

\section{Diagonalizing H}\label{section3}
\subsection{Reformulating the problem}
 We will now present an application of a diagonalization method which was developed further in \cite{mischa} . It works especially well with tridiagonal matrices like our $H$. For the reader's convenience we will explain the necessary details in the present paper as well. But let us first mention some general facts which will be useful later. 
 
 \begin{theorem}
 For any tridiagonal symmetric $(K+1)\times (K+1)$ matrix $M$ with coefficients $d_n, h_n$ labeled as in \ref{tridiag} holds:
 \begin{enumerate}
  \item \begin{equation}\label{detrel}
  \det(M_n)=d_{n-1} \det(M_{n-1})-h_{n-1}^2\det{M_{n-2}},
  \end{equation}
   where $M_n$ is the upper left $n\times n$ submatrix of $M$, the determinant of the $0\times 0$ matrix is $1$ and $n=1,\dots,K+1$.
  \item The eigenvalue equation $M v=E v$ is equivalent to the recursion relation 
  \begin{equation}\label{recrel}
  h_{n}v_{n-1}+h_{n+1}v_{n+1}+d_nv_n=E\, v_n
  \end{equation}
 with the initial and boundary conditions
 \begin{equation}\label{recboundary}
  v_{-1}=0 \text{ and } v_{K+1}=0,
 \end{equation}
where $v_j$ are the coefficients of $v$. Therefore, an eigenvector of $M$ is completely determined by the corresponding eigenvalue and by its first non-vanishing coefficient, which we can normalize to 1.
 \end{enumerate}
\end{theorem}
The second statement follows immediately from the eigenvalue equation; the first statement can be easily proved from the Laplace expansion of the determinant. Since our Hamiltonian becomes tridiagonal in the basis chosen above, both statements apply to it with $K$ replaced by $N/2$.\\

We can rewrite the matrices $H_B$ and $H_I$ in the following form:
\small
\begin{align}\label{finH}
 &H_B=-\frac{\alpha}{2}\sum\limits_{n=0}^{N/2} (n+1)\ketbra{n+1}{n}+n \ketbra{n-1}{n} +(2-\alpha) \sum n \ketbra{n}{n}  \nonumber\\
 &H_{I_1}=\frac{c}{N} \sum\limits_{n=0}^{N/2} n^2 \ketbra{n}{n},
\end{align}
\normalsize
where for brevity we have set $c:=\frac{3\alpha}{2}$.
We will not need $H_{I_2}$ in what comes, which is why we do not include it here.\\

In order to take a clean $N\rightarrow \infty$ limit we embed the matrices in an infinite dimensional Hilbert space, i.e. we view the finitely many basis vectors $\{ \ket{n} \}$ as subset of an infinite ONB of a function space. We will choose the Laguerre polynomials, because this ONB is particularly tailored to our needs. I.e. in the language of 1-dim. quantum mechanics we write:
\begin{equation}
 \braket{x}{n}=b_n(x):=\ee^{-\frac{1}{2}x}L_n(x).
\end{equation}
The functions $b_n(x)$ form an ONB of $L^2(0,\infty)$. Any finite or suitable infinite vector $v=(v_n)$ can be identified with a function/distribution via
\begin{equation}
 v(x)=\sum v_n\, b_n(x).
\end{equation}
The infinite dim. counterparts of $H_B$ and $H_{I_1}$ correspond to operators defined by \ref{finH} with the sums going to infinity instead of $N/2$. Let us call them $H_{B}^{\infty}$ and $H_{I_1}^{\infty}$, we will shortly show that they are differential operators. Along the lines of \cite{mischa} we want to use the solutions of the infinite dimensional eigenvalue problem 
\begin{equation}\label{diffeq}
 H^{\infty}v(x)=E\,v(x)
\end{equation}
 on $L^2(0,\infty)$
in order to find solutions of the finite dimensional problem by incorporating finite size effects. \\

The infinitely many coefficients of such a solution, i.e.
 $$v_n:=\braket{n}{v}_{L^2(0,\infty)}=\int\limits_0^{\infty}v(x)\,b_n(x)\, dx$$ 
will solve the recursion relation \ref{recrel}. However, they will not in general fulfill the boundary condition $v_{N/2+1}=0$. Those solutions of \ref{diffeq} that do fulfill $v_{N/2+1}=0$ (or $v_{N/2+1}\approx 0$) will be exact (or approximate) solutions to the finite dimensional problem.
\subsection{Finding the operators}
To make any use of the reformulation, we must find a not too difficult expression for the operators $H_{B}^{\infty}$ and $H_{I_1}^{\infty}$. Now it will become clear why we have chosen Laguerre polynomials as our basis - the infinite matrices take a particularly nice form expanded in them.
\vspace{0.5cm}

Laguerre polynomials fulfill two relations which will be important for us:
\begin{align}
        (n+1)&L_{n+1}(x)+n\;L_{n-1}(x)=(2n+1-x)L_n(x) \label{lag2}\\[3mm]
        &n\;L_n(x)=-(x\partial_x^2+(1-x)\partial_x)L_n(x).  \label{lag1}
\end{align}
From \ref{lag1} we can easily derive: 
\begin{equation}
 -\left(x\,\partial_x^2+\partial_x+\frac{2-x}{4}\right)\,b_n(x)=n\cdot b_n(x).
\end{equation}

With that we can evaluate the expressions appearing in $\bra{x}H^{\infty}\ket{y}$, the integral kernel of $H^{\infty}$, part by part.
\begin{enumerate}
 \item The matrix with $n$ on the diagonal and zero everywhere else corresponds to the integral kernel 
 \begin{align*}
        &\sum n\, b_n(x)b_n(y)=  -\left(x\,\partial_x^2+\partial_x+\frac{2-x}{4}\right)\, \sum b_n(x)b_n(y)\\
        &=  -\left(x\,\partial_x^2+\partial_x+\frac{2-x}{4}\right)\, \delta(x-y).
\end{align*} and hence to the operator
\begin{equation}\label{diagn}
 -\left(x\,\partial_x^2+\partial_x+\frac{2-x}{4}\right)
\end{equation}
\item The matrix with $n$ on the off-diagonal and zero everywhere else corresponds to the integral kernel
\begin{align*}
 &\sum \left((n+1)\, b_{n+1}(x)+n\, b_{n-1}(x)\right)b_n(y)=2\sum  n\, b_n(x)b_n(y)+(1-x)\delta(x-y),
\end{align*}
and hence to the operator
\begin{equation}
 -2\left(x\,\partial_x^2+\partial_x+\frac{2-x}{4}\right)+(1-x).
\end{equation}
We have used \ref{lag2} for the first equality and \ref{diagn} for the last step.
\item The matrix with $n^2$ on the diagonal corresponds to the fourth order differential operator 
\begin{equation}
  \left(x\,\partial_x^2+\partial_x+\frac{2-x}{4}\right)^2.
\end{equation}
Luckily, we will not need to solve equations involving this nasty operator since it is ``$1/N$ suppressed'' \footnote{One needs to be careful with this statement - certainly the spectrum of this operator is unbounded from above, as follows directly from its construction. We will be very precise in what sense the suppression plays a role.}  - instead we will be able to include its (extremely important) effects by developing a perturbation theory in $\frac{1}{N}$.
\end{enumerate}
In total we get:
\begin{align}
&H^{\infty}_B=-\frac{\alpha}{2}(1-x)-2(1-\alpha)\left(x\,\partial_x^2+\partial_x+\frac{2-x}{4}\right) \nonumber \\
&H_{I_1}^{\infty}=  \frac{c}{N} \left(x\,\partial_x^2+\partial_x+\frac{2-x}{4}\right)^2.
\end{align}
We can immediately notice the special role of $\alpha=1$, the critical coupling. As $\alpha$ crosses this value the differential part of $H_B^{\infty}$ changes its sign. Exactly at the critical value, $H_B^{\infty}$ is just a multiplication operator.  
\subsection{Diagonalizing $H_{B}$}\label{subsection31}
\subsubsection{Double scaling and breakdown of the Bogoliubov approximation}
Let us first diagonalize $H_B$ for $\alpha<1$ using its infinite counterpart. The infinite dim. eigenvalue equation
\begin{equation}
H_B^{\infty} \, v(x)=E\,v(x)
\end{equation}
can be solved exactly, e.g. using Mathematica. The (regular) solution for a fixed value $E$, let us call it $v^E(x)$, is:
\begin{equation}
v^E(x)=\ee^{-\frac{x}{2 \varepsilon }} L_{\frac{2E+(1-\varepsilon )^2}{4 \varepsilon }}\left(\frac{x}{\varepsilon }\right)
\end{equation}
with $\varepsilon:=\sqrt{1-\alpha}>0$. The solution $v^E(x)$ is in $L^2(0,\infty)$ iff the index of the Laguerre polynomial is a natural number. Therefore, the spectrum of $H_B^{\infty}$ is discrete, the eigenvalues are given by
\begin{equation}\label{infen}
\frac{2E_m+(1-\varepsilon )^2}{4 \varepsilon }=m \Longleftrightarrow E_m=2\varepsilon m-\frac{1}{2}(1-\varepsilon)^2.
\end{equation}
The corresponding eigenfunctions are
\begin{equation}\label{infvec}
v^m(x)=\ee^{-\frac{x}{2 \varepsilon }} L_{m}\left(\frac{x}{\varepsilon }\right)=b_m\left(\frac{x}{\varepsilon}\right),
\end{equation} 
i.e. just rescaled basis elements.
At this point one recognizes the familiar Bogoliubov spectrum. This should not be surprising - $H_B$ corresponds precisely to the operator 
$$
H_B\,\widehat{=} \, \varepsilon\left(b^\dagger_{1}b_1+b^\dagger_{-1}b_{-1}\right)
$$
with $b_k=u_ka_k+v^*_ka^{\dagger}_{-k}$ being the usual Bogoliubov modes, see e.g. \cite{nico}. Since we restrict ourselves to the $P=0$ sector, we expect the eigenstates to be related to $\left(b^{\dagger}_1b^{\dagger}_{-1}\right)^m\ket{0}_b$, which have precisely the energies $E_m$. It is tempting to just write $H_B\,= \, \varepsilon\left(b^\dagger_{1}b_1+b^\dagger_{-1}b_{-1}\right)$ as is often done in the literature. However, we prefer not to do so, since the Hilbert space on which $H_B$ acts is \textit{finite dimensional} and this finiteness has extremely important effects on the spectrum of $H_B$. In fact, these effects become dominant at the phase transition, i.e. at $\varepsilon=0$; also for $\varepsilon>0$ only a part of the spectrum of $H_B$ is given by $E_m$ and finite size effects must be taken into account. Therefore we leave $H_B$ as the matrix that it is and  explain how the finite $N$ effects set in carefully below. \\

Let us now see how the solutions \ref{infvec} encode the finite dimensional eigenvectors of $H_B$. The coefficients
\begin{equation}\label{coeff2}
v^m_n:=\braket{b_n(x)}{v^m(x)}=\int\limits_0^{\infty} b_n(x)\,b_m\left(\frac{x}{\varepsilon}\right) \,dx
\end{equation}
solve the recursion relation \ref{recrel}; but they need to fulfill the boundary condition $v^m_{N/2+1}\approx 0$ as well in order to correspond to real eigenvectors of $H_B$. We can calculate the following: 
\begin{equation}\label{coeff1}
v^m_n= \, \left(\frac{1-\varepsilon}{1+\varepsilon}\right)^n \cdot\frac{\varepsilon}{1+\varepsilon}\cdot \frac{P_{m}(\varepsilon,n)}{(1-\varepsilon^2)^m}
\end{equation}
where $P_{m}(\varepsilon,n)$ is a polynomial of order $2m$ in $\varepsilon$ and of order $m$ in $n$. For fixed $m$, the sequence of coefficients $(v^m_n)_{n\in\N}$ first oscillates for a while and then goes to zero exponentially fast. The bigger $m$ is and the smaller $\varepsilon$ is, the slower it goes to zero. We can summarize this in the following theorem:

\begin{theorem}
 \hspace{1mm}
\begin{enumerate}
 \item For every fixed $m\in \N$ and $1\geq\varepsilon>0$ the vector $(v_n^m)_{n\in\N}$ is localized up to $n_0(m,\varepsilon)\in\N$. The localization scale $n_0$ grows with increasing $m$ and decreasing $\varepsilon$. We have the limits 
 $$
 \lim\limits_{\varepsilon\rightarrow 0}n_0(m,\varepsilon)=\lim\limits_{m\rightarrow \infty}n_0(m,\varepsilon)=\infty \quad\text{and}\quad n_0(m,1)=m. 
 $$
 \item For fixed $N$ and $1\geq\varepsilon>0$ we define $m_{max}(N,\varepsilon)$ to be the biggest $m$ s.t. $n_0(m,\varepsilon)\leq N$. Since in equation \ref{coeff1} both $m$ and $n$ appear in the exponent, we observe that $m_{max}(N,\varepsilon)\approx \kappa_\varepsilon \cdot N$ to highest order in $N$, where  $\kappa_\varepsilon$ is an $\varepsilon$-dependent proportionality constant. \footnote{One can give a more careful argument to show that $m_{max}$ is proportional to $N$ to highest order. However, we do not want to concentrate too much on additional calculations, but rather present the necessary concepts as clear as possible.}
\end{enumerate}
\end{theorem} 

This theorem is quite powerful - it contains the Bogoliubov approximation and the double scaling limit. Indeed, for every fixed $N$ and $1\geq\varepsilon>0$, we find that the lowest $m_{max}(N,\varepsilon)$ eigenstates of $H_{B}$, let us call them $\ket{v^1},\dots,\ket{v^{m_{max}}}$, are well described by the expression \ref{coeff2} with the corresponding energy eigenvalues being $E_m$. Furthermore, for such states we have $H_I\ket{v^m}=\mathcal{O}\left(\frac{n_0^2(m,\varepsilon)}{N}\right)$. Since $m_{max}$ is proportional to $N$, we find that by letting $N$ go to infinity we can describe arbitrary many low lying energy eigenstates of $H$ with \ref{coeff2} to an arbitrary good approximation - as long as $n_0\ll\sqrt{N}$ we are on the safe side and the Bogoliubov approximation is valid. This statement can be made quantitative by finding an exact expression for $n_0$. For $m>m_{max}$ the vector given by \ref{coeff2} does not automatically satisfy the necessary boundary condition and hence does not even represent an eigenvector of $H_B$ (not to mention an eigenvector of $H$) - the finite size effects have kicked in. Then it might be possible to satisfy the boundary condition by tuning $m$ (while still keeping it a natural number), or, if that does not work, abandon the expression \ref{coeff2} completely. Indeed we see that the eigenvalues after $m>m_{max}$ differ from $E_m$, the spectrum of $H_B$ stops being linear at that point. That all of the above indeed holds can be read off the spectra of $H_B$ for different $\alpha$ and $N$, see the figures above.\\

One can also derive $m_{max}$ for the full interacting system in the usual Bogoliubov double scaling language. It is even possible to get an upper bound for the scaling of the energy gap with $N$ from the double scaling limit, by going as close as possible to the critical point s.t. the Bogoliubov approximation still holds for the lowest excited state. This was done in \cite{giamischa} and the upper bound calculated there turns out to give the correct scaling, see section \ref{engap}.

\begin{figure}
    \centering
    \begin{minipage}{0.5\textwidth}
        \centering
        \includegraphics[width=1.0\textwidth]{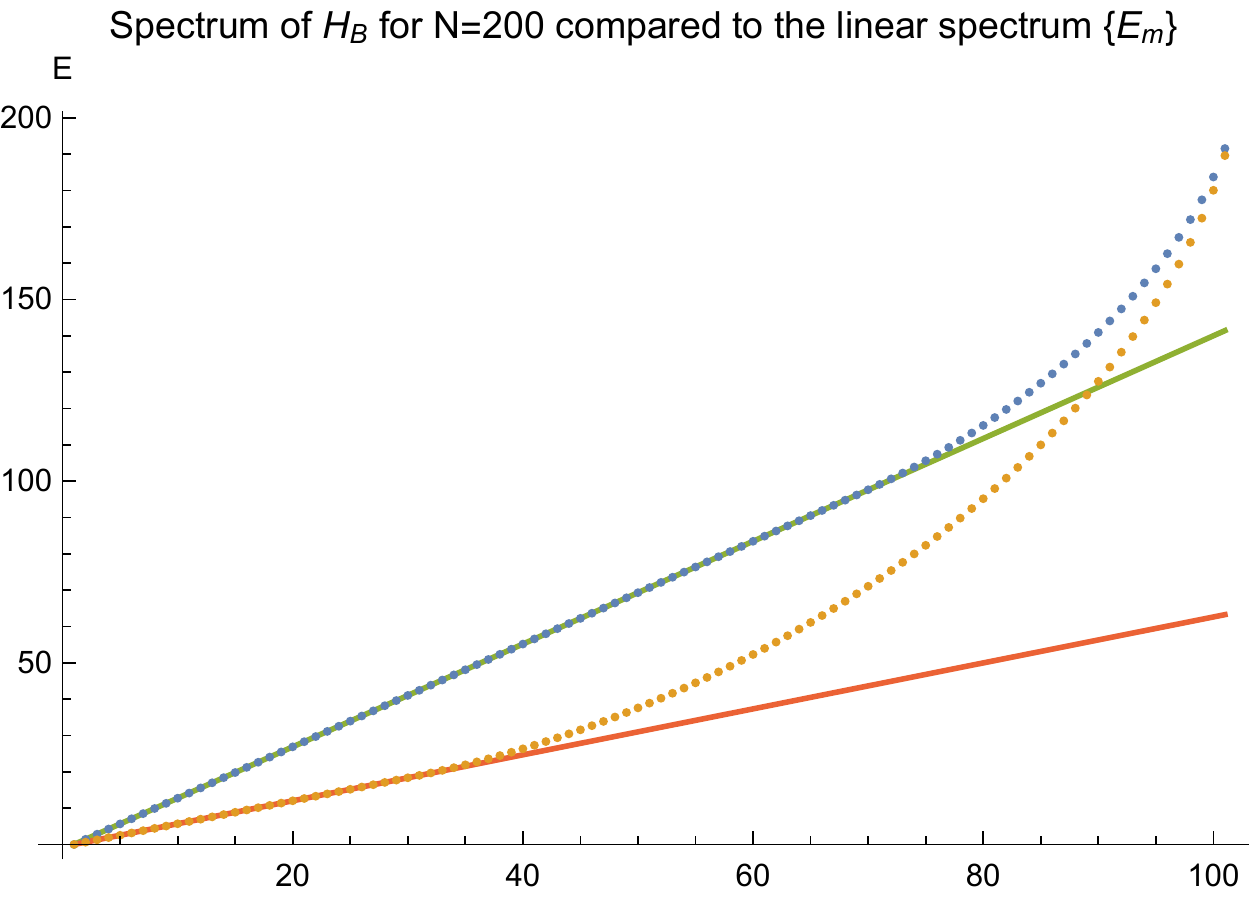}
    \end{minipage}%
    \begin{minipage}{0.5\textwidth}
        \centering
        \includegraphics[width=1.2\textwidth]{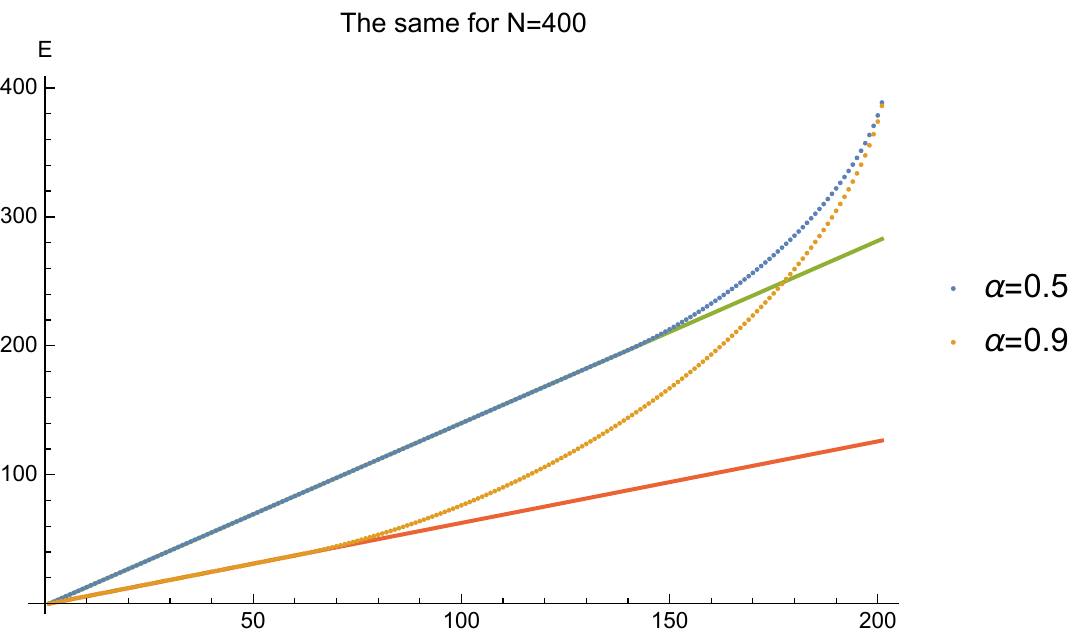}
         \end{minipage}
\end{figure}

\subsubsection{$H_B$ at the critical coupling}
If one diagonalizes first and then takes $\varepsilon\rightarrow0$, this limit appears very singular. The localization scale tends to infinity for every $m$ and the spectrum $\{E_m \mid m\in\N \}$ collapses to one point. However, it is perfectly fine to first set $\varepsilon$ to zero and then to diagonalize. In fact it is tremendously easy to find the spectrum of both $H_B^{\infty}$ and $H_B$ exactly at the critical point. Here is how it works:\\

At the phase transition, i.e. at $\alpha=1\Leftrightarrow \varepsilon=0$, the infinite dimensional version of the Bogoliubov Hamiltonian is simply a multiplication operator - we have $H_{B}^{\infty}=\frac{1}{2}(x-1)$. The infinite dim. eigenvalue equation becomes
\begin{equation}
\frac{1}{2}(x-1)\cdot v(x)=E\,v(x).
\end{equation} 
It might seem a bit unusual, but can be solved very easily. The solutions are delta functions labeled by the energy, we have:  
\begin{equation}
v^{E(y)}(x)=\delta(x-y)=\ee^{-x}\sum L_n(x)L_n(y)
\end{equation}
with the corresponding energies $E(y)=\frac{1}{2}(y-1)$. The coefficients $v^E_n$ can be read off the expansion of the delta, they are
\begin{equation}\label{evec}
v^{E(y)}_n=L_n(y)=L_n(2E+1).
\end{equation}
The spectrum is continuous, and the eigenfunctions are no longer localized nor normalizable; nevertheless\footnote{ or, better said, precisely due to the continuity of the spectrum} the finite size effects can be included \textit{exactly} by finetuning $E$.\\

The boundary condition \ref{recboundary} becomes $L_{N/2+1}(2E+1)=0$. From that we conclude: Let $y_0,\dots,y_{N/2}$ be the $N/2+1$ roots of the Laguerre polynomial $L_{N/2+1}$. Then the eigenvalues of $H_B$ are given by $E_m=\frac{1}{2}(y_m-1)$, and the coefficients of corresponding eigenvectors (labeled by n) are $v^{E_m}_n=L_n(y_m)=L_n(2E_m+1)$. Exactly at the phase transition, the boundary condition is the dominant effect!\\
It is not hard to see directly that these results are exact:
\begin{theorem}
The characteristic function of the matrix $H_{B}$ at the phase transition is 
\begin{equation}
\det(H_{B}-E\cdot\mathbb{1})=\frac{(N/2+2)!}{2^{N/2+2}}L_{N/2+1}(2E+1).
\end{equation}
\end{theorem}
To prove this, one plugs the relation in the recursion relation for the determinant of tridiagonal matrices \ref{detrel}, and use the recursion relation for the Laguerre polynomials \ref{lag2} to conclude the result. 

Also, it is easy to see that the expression for the eigenvectors \ref{evec} indeed solves the corresponding recursion relation \ref{recrel} with the correct boundary condition.

\subsection{Including the interaction}\label{subsectionint}
\subsubsection{Overview of the problem}
In the previous section, we saw how for any fixed $\varepsilon>0$ the Bogoliubov approximation is valid in the ``double scaling'' limit as $N$ goes to infinity. We have also seen how the approximation fails at the critical point $\varepsilon=0$ - every eigenvector of $H_B$ feels the finiteness of the matrix and hence is completely delocalized. Therefore, right at the critical point, no single eigenvector of $H_B$ approximates an eigenvector of $H$.\\
In this section, we will find the low energy eigenvectors of $H_B+H_{I_1}$ to first order in $1/N$. Although a priori it might be that including just $H_{I_1}$ is not enough, it turns out that it is. This is because the low energy eigenstates of $H_B+H_{I_1}$ are localized up to $n_0\sim N^{1/3}$, so that the contribution of $H_{I_2}$ to those vectors is $\mathcal{O}\left(N^{-\frac{1}{6}}\frac{N^{\frac{2}{3}}}{N}\right)=\mathcal{O}\left(N^{-\frac{1}{2}}\right)$ and therefore is negligible for large $N$ \footnote{The additional factor $N^{-\frac{1}{6}}$ comes from the normalization of the eigenstates}.\\

To appreciate the difficulty of the problem, it is useful to notice two things. First, we will obtain an explicit expression for the low energy states of an interacting many body system at the critical point, i.e. we will need to work beyond any mean field approximation. Second, the infinite dimensional version of $H_B+H_{I_1}$ is a fourth order differential operator, namely
$$
H_B^{\infty}+H_{I_1}^{\infty}= \frac{1}{2}(x-1)+\frac{3}{2N} \left(x\,\partial_x^2+\partial_x+\frac{2-x}{4}\right)^2,
$$
and we will find approximate eigenfunctions of this operator. Obviously, diagonalizing $H_B+H_{I_1}$ is a non-trivial task and we have to come up with a smart way to perform a $1/N$ expansion.\\

The idea is to use the information that we obtained about the spectrum of $H_B$ and to perturb around it. So we are after something like a Rayleigh-Schroedinger perturbation theory. That Rayleigh-Schroedinger itself cannot work is clear from the start - $H_{I_1}$ is not a small perturbation and the eigenstates of $H_B+H_{I_1}$ differ drastically from the eigenstates of $H_B$. Instead, we will develop what we would like to call \textit{a perturbation theory on the level of recursion relations}. So, we will use the formulation \ref{recrel} and \ref{recboundary} of the eigenvalue problem.

\subsubsection{Finding an expression for the eigenstates}

The eigenvalue equation $ (H_B+H_{I_1})\cdot \ket{v} = E\,\ket{v}$ is equivalent to the recursion relation 
\begin{equation}\label{recrel2}
 h_{B;\,n+1}v_{n+1}+h_{B;\,n}v_{n-1}+(d_{B;\,n}+d_{I_1;\,n})v_n=E\, v_n
\end{equation}
(no summation over n) with the initial values $v_{-1}=0, v_0=1$ (because we know that the low energy states have an overlap with the condensate state) and the boundary condition $v_{N/2+1}=0$. In this language, including the interaction is equivalent to changing $d_n\mapsto d_n+\frac{3n^2}{2N}$. We already know the solution of relation \ref{recrel2} without the interaction; let us call it $v_{B;\,n}(y):=L_n(y)$ with $y=2E+1$ as above. Now let $v_{full;\,n}(y)$ be the solution to the ``interacting recursion" and let us split it as 
\begin{equation}\label{deltav}
 v_{full;\,n}(y)=:v_{B;\,n}(y)+ v_{I;\,n}(y).
\end{equation}
We want to understand how $v_{I;\,n}$ behaves in the region $n\ll N$. For that we insert \ref{deltav} into the interacting recursion relation and ignore the terms $\frac{n^2}{N}v_{I;\,n},$ as these are additionally suppressed. We end up with the following equation for $v_{I;\,n}$:
\begin{equation}\label{intrec}
 h_{n+1}v_{I;\,n=1}+h_{n}v_{I;\,n-1}+d_{n}v_{I;\,n}=E\, v_{I;\,n}-\frac{3n^2}{2N}v_{B;\,n}(y), 
\end{equation}
with initial values $v_{I;\,0}=0, v_{I;\,1}=\frac{3}{2N}$ (we have used the explicit form of $h$, $d$ and $v_0$ to get the expression for $v_{1\,I}$). This is like solving an inhomogeneous system for which the homogeneous solutions are known - it can be done! Indeed, e.g. with Mathematica, one can obtain an explicit analytic solution of the relation \ref{intrec} for any inhomogeneous term. The result contains messy expressions, but schematically it is simple:
\small
\begin{align}\label{vint}
 &v_{I;n}(y)=\frac{3}{2N}\cdot \\
 & \left(\gamma_1(n,y) \sum _{m=0}^{n} m^2 v_{B;\,m}(y) \beta_1(m,y) + \gamma_2(n,y) \sum _{m=0}^{n} m^2 v_{B;\,m}(y) \beta_2(m,y) - \frac{\gamma_2(n,y)}{y-1} \right),\nonumber 
\end{align}
\normalsize
where $\gamma_1,\gamma_2$ are the two homogeneous solutions of the recursion relation, both of order one for small $y$. The coefficients $\beta_1,\beta_2$ are rather complicated sums and products of Laguerre polynomials, but the only thing important for us is that they are all of order one for small $y$, i.e. for low energies. From that we can immediately conclude that $v_{I;n}(y)$ grows as $\frac{n^3}{N}$, and this is all we need for the following discussion.\\

Let us reflect shortly on what we did above. First, it is clear how to get to higher orders in $1/N$: one includes the knowledge of the first order solution, let us call it $v_{full,\,n}^1(y)$ and splits 
\begin{equation}
 v_{full;\,n}(y)=:v_{full,\,n}^1(y)+v_{full,\,n}^2(y).
\end{equation}
After inserting this into the recursion relation \ref{recrel2} and ignoring the terms $\frac{n^2}{N}v_{full,\,n}^1(y)$, one gets a relation similar to \ref{intrec} for $v_{full,\,n}^2$. In other words the higher order correction can be obtained exactly as in the Rayleigh-Schroedinger perturbation theory. The advantage of working with recursion relations instead of with matrices, is that for the former it is easier to identify the region wherein the perturbation is small and to stay in it, whereas for the latter this is not so straightforward.

\subsubsection{The localization scale and the energy gap}\label{engap}
Let us now try to understand what expression  \ref{vint} can tell us about the eigenvectors. It is useful to have a look at the numerical results to get a feeling for what we are looking for. One observes that the low energy eigenstates of $H_B+H_{I_1}$ are localized up to some $n_0(N)$, which grows with $N$, but for which $n_0^2/N$ decreases, i.e. for large enough $N$ the localization scale $n_0$ should fall within the region where our approximations are valid. We are lucky. \\
Now notice the following: whenever two subsequent coefficients, say $v_{n_0}, v_{n_0+1}$, of a solution of the recursion relation of the type \ref{recrel} are both zero, the recursion stops and all following coefficients $v_{n_0+k}$ can be consistently set to zero. That is, for real eigenvectors of  $H_B+H_{I_1}$ the corresponding energy eigenvalue is so to speak \textit{finetuned}, so that at some point $n_0$, two subsequent coefficients are zero and the eigenvector is localized. We cannot hope to see this finetuning in our approximated expression for $v_{full;\, n}$, but we can incorporate its effects by hand.\\

Above, we have split the coefficients $ v_{full;\,n}(y)=v_{B;\,n}(y)+ v_{I;\,n}(y)$ and we know that $v_{B;\,n}(y)$ is completely delocalized. So it is a cancellation between $v_{B;\,n}(y)$  and $v_{I;\,n}(y)$ that leads to the localization of $v_{full}$. This cancellation can only happen when both parts of $v_{full}$ are of the the same order - and since $v_{B;\,n}(y)$ stays (on average) of order one while $v_{I;\,n}(y)$ grows as $\frac{n^3}{N}$, we find that the localization scale $n_0$ must be of order $n_0\sim N^{\frac{1}{3}}$. This indeed coincides with the numerical results. Furthermore, in order to obtain an approximate expression for the real eigenvector $v_{full}$, we must cut the approximate analytic result obtained from \ref{vint} whenever two subsequent coefficients  $v_{n_0}, v_{n_0+1}\approx 0$  while $n_0$ is of order $N^{1/3}$. This is because whenever  $v_{n_0}, v_{n_0+1}$ are not \textit{exactly zero}, the following coefficients $v_{n_0+k}$ continue to grow and hence start to deviate more and more from the real eigenvector. In the figure below for $N=40000$ the real ground state and first excited state (computed numerically) are plotted next to the first order analytic expression for $v_{full}$ and to the homogeneous solution $v_B(y)$. The difference between $v_{full}(y)$ and the real eigenvectors $v_{real}$ is 0.5\% for the ground state and 2.5\% for the excited one \footnote{ The difference is computed as $\frac{\norm{v_{full}-v_{real}}^2}{\norm{v_{real}}^2}$}.\\

\begin{figure}[ht]
	\centering
  \includegraphics[width=1.1\textwidth]{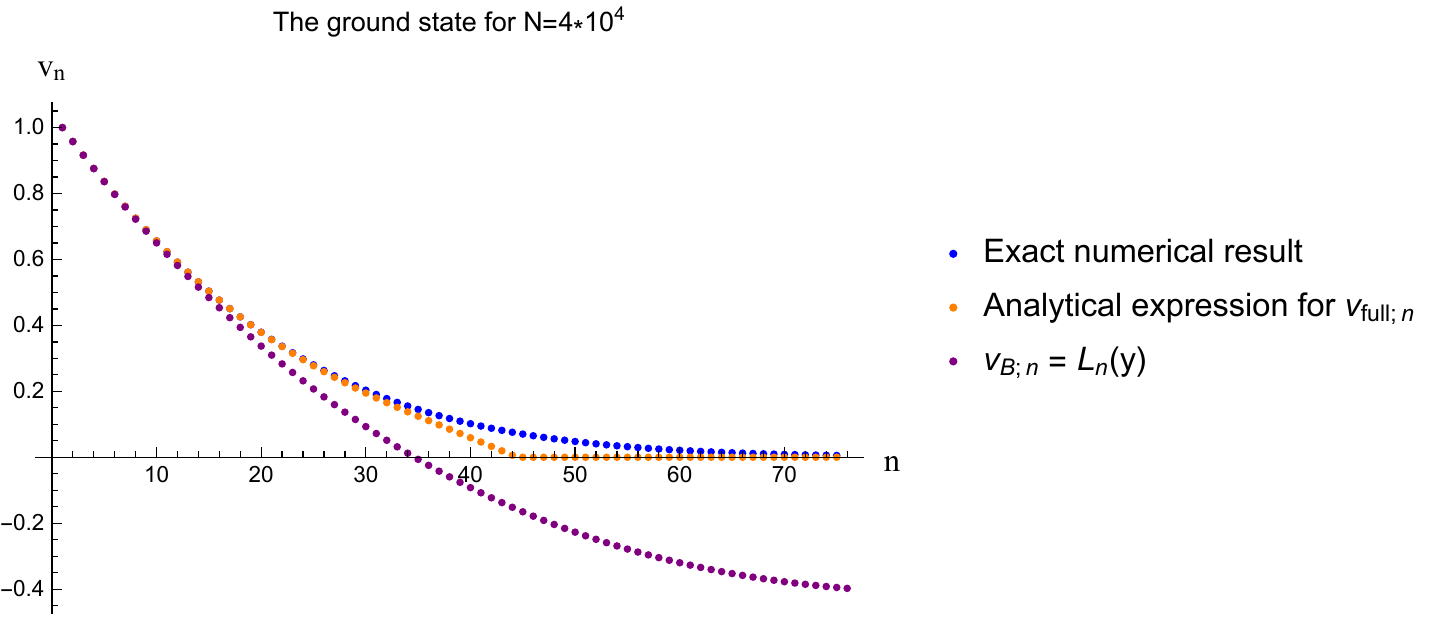}
		\label{gstate}
\end{figure}
\begin{figure}[ht]
	\centering
  \includegraphics[width=1.1\textwidth]{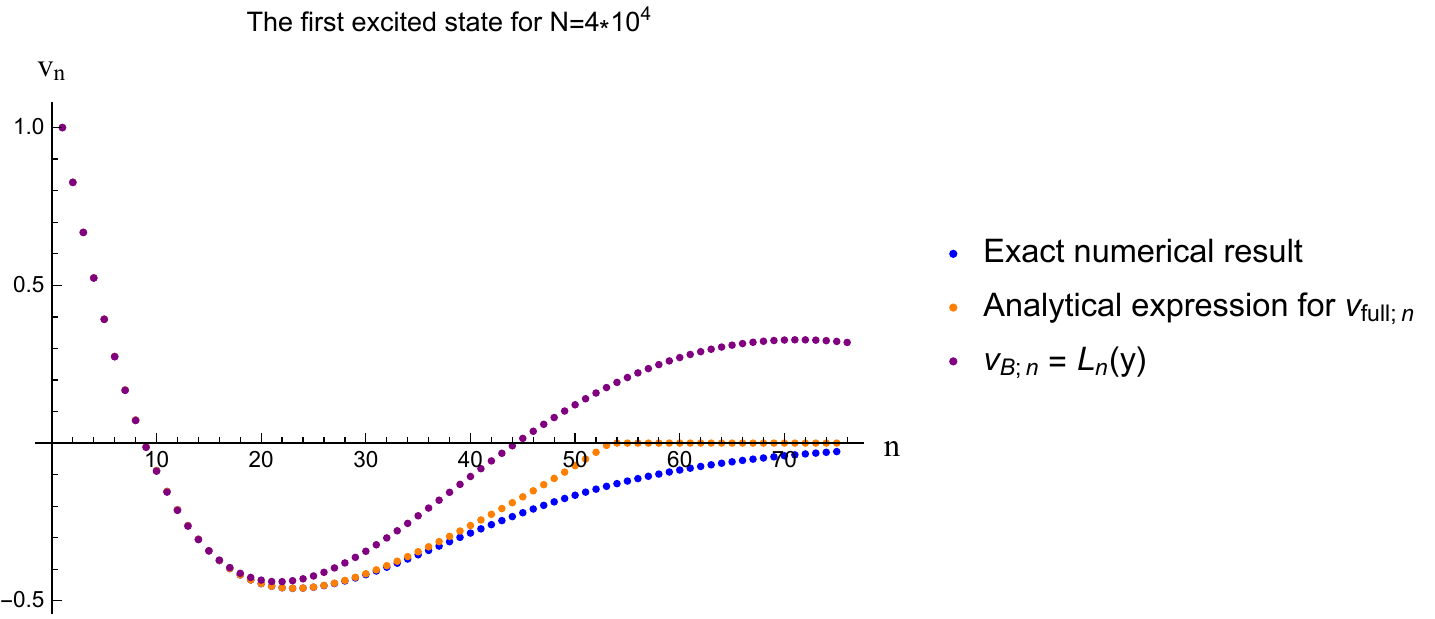}
	
	\label{fstate}
\end{figure}

We can also infer the scaling of the energy gap between the low energy eigenvalues from \ref{vint}. For that we notice that $y$ enters expression \ref{vint} only through $L_n(y)$, and for small energies we can write $L_n(y)\approx 1-n\,y$. The cancellation between $v_{B;\,n}(y)$  and $v_{I;\,n}(y)$ happens at $n\sim N^{1/3}$, so effectively it is $N^{1/3}y$ that must be finetuned. From this we conclude that the energy gap is of order $N^{-\frac{1}{3}}$ for the low energy eigenstates, since this is the characteristic scale for $y=2E+1$. So at the critical coupling the system becomes gapless as $N^{-1/3}$ - this gaplessness in the infinite $N$ limit (though not the explicit scaling behavior) was also observed e.g. in \cite{nico} and \cite{gold} using very different methods. This scaling of the energy gap coincides with numerical results to a surprising accuracy, see the image below.\\

\begin{figure}[ht]
	\centering
  \includegraphics[width=1.1\textwidth]{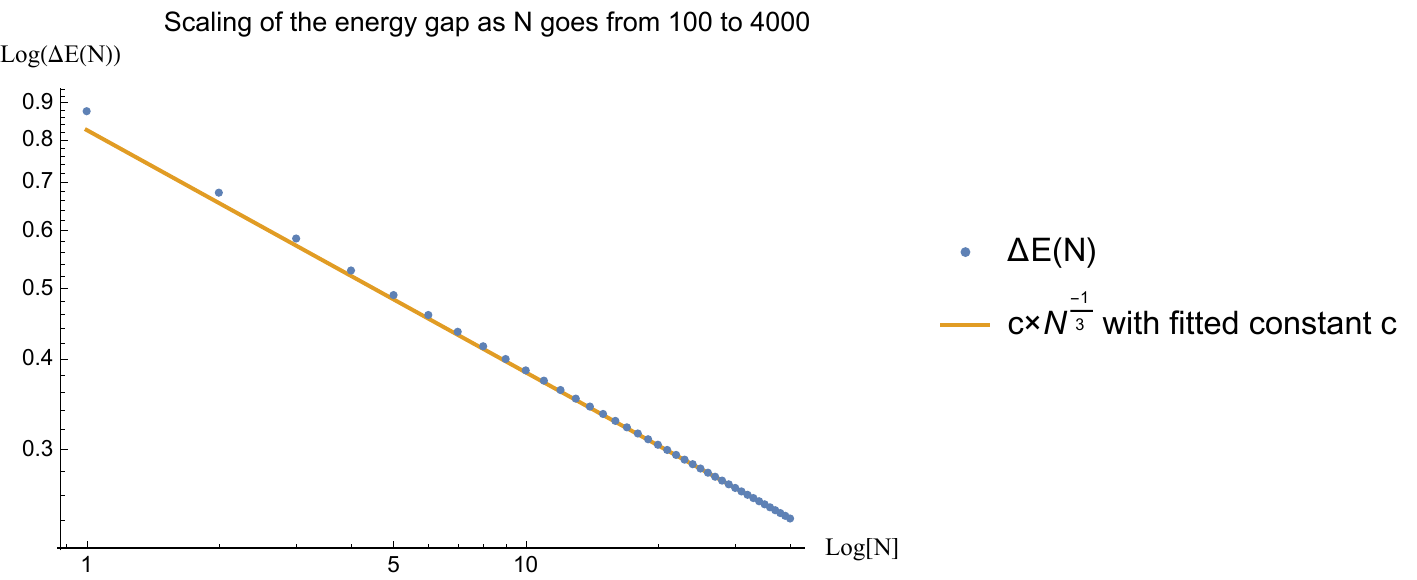}
	\label{scaling}
\end{figure}

\section{Contribution of the high momentum modes}\label{sectionmodes}
We will now show that the higher momentum modes decouple from the low energy sector as $N$ goes to infinity, which is why our calculation makes sense in the first place. The crucial reason is the localization scale $n_0\sim N^{\frac{1}{3}}$ and the scaling of the energy gap  $\Delta E\sim 1/n_0$ of the 3-mode matrix.\\

Let us restrict ourselves to the first $\Lambda$ modes, fix the particle number to be $N$ and the total momentum to be zero. The Hilbert space is finite dimensional, it is spanned by the basis elements
\begin{equation}
 \ket{n_1,\dots,n_{\Lambda}} \text{ restricted to } \sum n_j \leq \frac{N}{2},
\end{equation}
where as in \ref{basis} $n_j$ are the occupation numbers of the $j$-th momentum mode and we have $n_j=n_{-j}$ and $n_0=N-2\sum n_j$. Let us forget about the restriction $ \sum n_j \leq \frac{N}{2}$ for the moment - it is a finite size effect and does not affect the following discussion. If we ignore it, we can write the Hilbert space as $(\R^{\frac{N}{2}+1})^{\otimes\Lambda}$ and the Hamiltonian as the following matrix:
\begin{equation}
 H=\left(H_B+H_{I_1}\right)\otimes\bigotimes\limits_{k=2}^{\Lambda}H_{B_k}+\widetilde{H}_I,
\end{equation}
where $H_B$, $H_{I_1}$ is as above, $H_{B_k}$ is a tridiagonal matrix of the type \ref{tridiag} with the entries $h_{n;\,B_k}=-\frac{\alpha}{2}n$, $d_{n;\,B_k}=(2k^2-\alpha)n$ and $\widetilde{H}_I$ is whatever is left of the Hamiltonian \footnote{ It is not just a tensor product of $\Lambda$ matrices}.  For vectors localized up to $\ket{n_1,\dots,n_{\Lambda}}$ we have $\widetilde{H}_I\ket{n_1,\dots,n_{\Lambda}}=\mathcal{O}\left( \frac{\max(n_k)^2}{N} \right)$. From this we can conclude that the low energy eigenstates of $H_B+H_{I_1}$ approximate true eigenstates of the full Hamiltonian for large $N$. We can also observe that \textit{only they} contribute to the low energy part of the spectrum. This is because the low energy states of each $H_{B_k}$ with $k>1$ is given by the (accordingly modified) expression \ref{coeff2} - the Bogoliubov transformation is not singular for these modes. Therefore the low energy eigenvalues coming from the inclusion of higher modes tend to $E_{m_k}:=2\varepsilon_k m-\frac{1}{2}(1-\varepsilon_k)^2$ and do not scale with increasing $N$. So the band of energy eigenvalues coming together with $\Delta E\sim N^{-\frac{1}{3}}$ is entirely governed by the 3-mode Hamiltonian \ref{tridiag}.

\section{The quartic oscillator}\label{sectionosc}
In \cite{oscillator} it was shown that for large $N$ the three mode Hamiltonian \ref{tridiag} corresponds to a particle in a two dimensional mexican hat potential. This was achieved by writing the momentum operators in the continuous description, i.e. $a_1=\frac{1}{\sqrt{2}}(x_1+\iu p_1), a_{-1}=\frac{1}{\sqrt{2}}(x_{-1}+\iu p_{-1})$  and neglecting the suppressed terms. One obtains (after a redefinition of the continuous variables $x_1,x_{-1}$):
\begin{equation}
V(r)=\frac{1-\alpha}{2}r^2+\frac{7\alpha}{32N}r^4 \quad \text{ with } r^2=x_1^2+x_{-1}^2
\end{equation}
 Precisely at the critical point the potential becomes purely quartic - the scaling of the energy gap can be rederived from an analysis of quartic oscillators, see e.g. \cite{oscillator},\cite{engap}. Thus the methods outlined above can be used to find eigenstates of the quartic oscillator perturbatively in the anharmonicity parameter. 

\section{Conclusions and Outlook}
Using the presented diagonalization method several properties of the Lieb Liniger model at the quantum phase transition were derived by means of simple calculations. We can observe the appearance of gapless states, find the scaling of the energy gap and the depletion with the particle number etc. We hope, that not only can this particular system be used (possibly even in laboratories) as a toy model for black holes, but also that the presented method will be useful in other contexts. Essentially, whenever one encounters a large sparse matrix, as is often the case in condensed matter systems, the method could potentially be useful. For now, we leave the extension of it to other systems for future work.

\section*{Acknowledgements}

First and foremost I would like to thank Gia Dvali for his support and encouragement during the preparation of this paper. Further, it is a pleasure to thank Daniel Flassig, Andre Franca, Peter Pickl and Nico Wintergerst for inspiring discussions and ongoing collaboration.
 
The work of Mischa Panchenko was supported by the ERC Advanced Grant ``UV-completion through Bose-Einstein Condensation" (Grant No. 339169).

\end{document}